\documentclass[conference]{IEEEtran}
\usepackage{graphicx}
\usepackage{subfigure} 
\usepackage{algorithm}
\usepackage{algorithmic}
\usepackage{bm}
\usepackage{multirow}
\usepackage{amsmath}
\usepackage{booktabs}
\usepackage{diagbox}
\usepackage{changepage}
\usepackage{enumerate}
\usepackage{subfigure}
\usepackage{caption}
\usepackage{booktabs}
\usepackage{soul}
\usepackage{hyperref}

\begin{document}

\title{Generative Poisoning Attack Method Against Neural Networks}

\author{
\IEEEauthorblockN{Chaofei Yang}
\IEEEauthorblockA{University of Pittsburgh\\
Pittsburgh, PA 15261\\
chy61@pitt.edu}
\and
\IEEEauthorblockN{Qing Wu}
\IEEEauthorblockA{Air Force Research Laboratory\\
Rome, NY 13441\\
qing.wu.2@us.af.mil}
\and
\IEEEauthorblockN{Hai Li, Yiran Chen}
\IEEEauthorblockA{Duke University\\
	Durham, NC 27708\\
	{hai.li, yiran.chen}@pitt.edu}
}

\maketitle

\begin{abstract}
	\label{sec:abstract}
	Poisoning attack is identified as a severe security threat to machine learning algorithms.
	%Poisoning attack has been a severe issue to machine learning algorithms.
	In many applications, for example, deep neural network (DNN) models collect public data as the inputs to perform re-training, where the input data can be poisoned.
	Although poisoning attack against support vector machines (SVM) has been extensively studied before,
	there is still very limited knowledge about how such attack can be implemented on neural networks (NN), especially DNNs.
	%In the past, the SVM-oriented poisoning attack techniques have been extensively studied.
	%While it lacks of systematic exploration on how to attack the emerging machine learning algorithm---neural networks, especially deep neural networks.
	%Very importantly, many new machine learning hardware and facilities, such as parallel distributed learning systems, cloud computing, and novel neuromorphic computing hardwares, are exposed to high threat risks of poisoning attacks as they often collect data from public and perform re-training.
	%
	In this work, we first examine the possibility of applying traditional gradient-based method (named as the direct gradient method) to generate poisoned data against NNs by leveraging the gradient of the target model w.r.t. the normal data.
	%we propose two poisoning attack schemes against NN.
	%The first approach leverages the gradient of normal data loss w.r.t. the poisoned inputs while
	We then propose a generative method to accelerate the generation rate of the poisoned data:
	%the second approach is based on a more sophisticated generative method:
	an auto-encoder (\textit{generator}) used to generate poisoned data is updated by a reward function of the loss, and the target NN model (\textit{discriminator}) receives the poisoned data to calculate the loss w.r.t. the normal data.
	Our experiment results show that the generative method can speed up the poisoned data generation rate by up to $239.38\times$ compared with the direct gradient method, with slightly lower model accuracy degradation.
	%demonstrate the effectiveness and efficiency of our proposed techniques.
	A countermeasure is also designed to detect such poisoning attack methods by checking the loss of the target model.
	%Experiment results demonstrate the effectiveness and efficiency of our proposed techniques.
	%The speedup of the generative method is up to $239.38\times$, comparing with the direct gradient method.
\end{abstract} 
\section{Introduction}
\label{sec:introduction}
Machine learning has been widely used in information processing to help users understand the underlying property of the data.
The example applications include image classification and recognition, feature extraction, language processing, video analysis and etc.
As one of the major types of machine learning models, NN processes input data by multiplying them with layers of weighted connections.
NNs have been deployed to a large variety of embedded systems to offer machine intelligence, where the NNs need to collect input data to perform re-training. For example, a small-footprint large vocabulary speech recognizer for mobile devices~\cite{lei2013accurate}, a small robust deep-learning model designed to provide high quality text-to-speech functionality on smart devices~\cite{borocs2015robust}, and an audio sensing framework built from coupled DNNs~\cite{canziani2016analysis}.
%Many embedded hardware engines have been developed to implement neurals network with high speed and efficiency, e.g., Neurogrid from Stanford University~\cite{benjamin2014neurogrid},  Human Brain Project~\cite{markram2012human}, TrueNorth from IBM~\cite{akopyan2015truenorth}, and BRAIN Initiative~\cite{insel2013nih}.
Such applications, however, introduce at least two types of security issues -- \textit{causative attack} and \textit{exploratory attack}~\cite{barreno2010security,huang2011adversarial}.

Exploratory attack does not change the parameters of the target model.
Instead, the attacker sends new data to the target model and observes the model's decisions on these carefully crafted input data.
One recent example of exploratory attack is \textit{adversarial example}~\cite{szegedy2013intriguing, goodfellow2014explaining}.
%hich is a counter-intuitive phenomenon called adversarial examples is identified and studied
The authors show that DNN can be fooled by adding small perturbations onto the inputs even though these perturbed inputs can still be easily recognized by humans.
%It depicts the truth that machine learning model has not learned the fundamental nature of the task.
No parameters of the NN are changed during the attack.
%An attacker can generate such examples for compromising a neural network, without changing parameters of it~\cite{goodfellow2014explaining}.
Another recently proposed exploratory attack scheme leverages the model's transfer learning ability to replicate the target model without the prior knowledge of its parameters and structure~\cite{yang2016securityofnc}.
%These attacks draw dramatical attention from deep learning society.
%Note that the aforementioned attacks don't change the parameters of the target model, therefore, they are classified as exploratory attacks. This depicts the attacker's capability of accessing the training data.

On the contrary, causative attack allows attackers to manipulate training dataset in order to change the parameters of the target model and reconstruct it.
One important type of causative attack is \textit{poisoning attack} where artificially poisoned data are sent into the target model with attacking labels.
The target model then updates itself with the poisoned data and gradually compromises.
%Poisoning attack is totally unique from adversarial attack, with different capabilities, generation algorithms, and application scenarios.
Several schemes have been proposed to conduct poisoning attack against SVMs~\cite{biggio2012poisoning,xiao2012adversarial}.
However, we have not seen many works about poisoning attacks against NNs.
%Unfortunately, there is rarely related work on neural networks of such attack.
One possible reason may come from the vague mathematic understanding of NN.
In a NN, for example, there is no explicit direct expression of gradient calculation as the gradients are updated during backpropagation.
This fact prevents gradient-based poisoning attack from being deployed in NNs where the second partial derivative needs to be calculated.
%This is because of the black box nature of neural networks and lack of theoretical support.
%The backpropagation algorithm makes the gradient calculation extremely difficult since there is no explicit expression.
%However, the growing of neuromorphic computing chips, parallel distributed learning system, and cloud computing is making them face potential threats of poisoning attacks as they usually need to be re-trained during operation or to collect data from the public. Therefore, we are in great need of preventing neural network model from such attack.

In this work, we first examine the possibility of applying traditional gradient-based direct gradient method to generate the poisoned data against NN.
We then propose a generative method to speed up the generation rate of the poisoned data by bypassing the gradient calculation, which is the bottleneck in the direct gradient method.
The proposed scheme is partially inspired by the concept of \textit{Generative Adversarial Network} (GAN)~\cite{goodfellow2014generative}:
An autoencoder (\textit{generator}) is used to generate the poisoned data and updated by a reward function of loss, and then sends the poisoned data to a discriminator.
The target NN model (\textit{discriminator}) receives the poisoned data and calculates the loss w.r.t. the normal data, and then sends the calculated gradients back to the generator.
We also design a reward function and a strategy of sending back the gradients.
%is used to process the loss and a strategy of sending back the gradient is designed.
Different from a traditional method that periodically checks the accuracy of the model~\cite{mozaffari2015systematic}, we also propose a loss-based method to detect poisoning attack with much lower computation overhead.
The proposed countermeasure is based on the fact that a normal input only introduces a small loss since it is close to the original decision region,
while a poisoned input usually incurs a large loss as it is far away from this region.
Hence, we are able to distinguish the types of the input by checking this loss difference.

Our major contributions can be summarized as:
\begin{itemize}
	\item We examine traditional gradient-based poisoning attack on NN and identify the poisoned data generation rate as the bottleneck of its implementation;
	\item Based on our examination, we propose a generative method to substantially speed up the poisoned data generation rate with slightly degraded model attack effectiveness, i.e., target model accuracy degradation;
	\item We proposed a loss-based countermeasure to detect the poisoning attack with very minimum overhead;
	\item The effectiveness of the proposed schemes are extensively evaluated by performing experiments on MNIST and CIFAR-10 datasets under different configurations.
\end{itemize}
%(i) we analyze the security threat brought by the poisoning attacks, specifically on NN-based machine learning algorithms;
%(ii) based on the analysis, we propose two methods to generate poisoned data against NNs for the first time;
%(iii) the effectiveness of these techniques is carefully evaluated by performing experiments on MNIST and CIFAR-10 datasets under different configurations;
%(iv) a loss-based countermeasure is proposed for poisoning attack detection, leveraging a distinct loss difference.

%The remainder of the paper is organized as follows:
%Section~\ref{sec:related} summarizes the related studies on poisoning attack;
%Section~\ref{sec:algorithm} first introduces the threat model of poisoning attack against NNs. It then describes implementations of the direct gradient method and the generative method. The proposed countermeasure is also presented in this section;
%%Our proposed threat model of poisoning attack against NNs and the techniques of generating poisoned data are described and explained in We also propose a countermeasure against such attacks in this section.
%Section~\ref{sec:experiment} shows our experimental results and discussions;
%%The experimental results and the comparison under different configurations are presented in Section~\ref{sec:experiment}.
%%At the end, we conclude the paper in Section~\ref{sec:conclusion}.
%Section~\ref{sec:conclusion} concludes this work.

\section{Related Work}
\label{sec:related}

The security attacks against learning algorithms can be mainly categorized into two types: exploratory attack (exploitation of the classifier) and causative attack (manipulation of training data)~\cite{barreno2010security,huang2011adversarial}
%depict that attacks against learning algorithms can be classified into causative (manipulation of training data) and exploratory (exploitation of the classifier). 
Poisoning attack belongs to the latter type and has been investigated in many prior arts.

One of the earliest studies on poisoning attack was performed on design of anomaly detector~\cite{rubinstein2009antidote}. The author showed how an attacker can substantially increase his chance to evade the detection by only introducing moderate amounts of poisoned data to the detector.
A PCA-based detector was also proposed based on the fact that poisoning attack significantly distorts the model produced by the original PCA method. 
%method was also proposed to anomaly detection, based on robust statistics since poisoning has little effect on the robust model.
A similar work was presented in~\cite{kloft2010online}.

There are many existing works about poisoning attack against SVM, thanks to its clear mathematic basis. 
%Most of the previous works on poisoning attack are against SVM, benefiting from its mature theoretical basis. 
In~\cite{biggio2012poisoning}, the author proposed a gradient ascent strategy in which the gradients are computed based on properties of the SVM's optimal solution. 
%Gradient calculation is always a workable method for the optimization problem. The training of neural networks is also based on gradient backpropagation. 
In learning algorithms, gradient calculation is the basis of many optimization problems, such as the training of NNs using gradient backpropagation.
%We take advantage of this idea in this paper. 
It also inspired the schemes proposed in our work.
In~\cite{xiao2012adversarial}, the author studied optimizations of label flipping attack -- another version of poisoning attack. 
Label flipping attack is interpreted as a linear programming problem, which can be solved by approximating the original problem with a simpler problem.
%The author interpreted such attack into linear programming problem and dealt with related techniques. 
%The idea of approximating the harder original problem with a proper simpler problem is helpful. 
Although the demonstrated solutions are effective, their applications may be limited due to the specific target model. %These works are inspiring and effective, however, their exclusive target model limits their application.

A systematic, algorithm-independent poisoning attack approach was presented in~\cite{mozaffari2015systematic}.
The proposed attack procedure is able to generate the poisoned input data based on only the distribution of training data. 
However, the author gives neither the proof of why the method works nor how to ensure the effectiveness of the proposed approach.
%and the method without consideration of the target model is hardly effective. 
%Building on ideas from these many previous works, we examine a gradient-based poisoning attack algorithm against NNs and further propose a generative method of poisoned data generation for the first time.
\section{Poisoning Attack Against Neural Networks}
\label{sec:algorithm}

\subsection{Notations}

In this manuscript, we use $\bm{x}_i$ and $\bm{x}_p$ to denote the normal data and the poisoned data, respectively.
Here, the subscripts $i$ and $p$ indicate the variables of the normal data and the poisoned data, respectively.
We also use superscripts $(o)$ and $(p)$ to indicate the variables of the original model and the poisoned model, respectively.
For example, $L_i^{(p)}(t)$ is the loss of the poisoned model w.r.t. the normal data, where $t$ is the time stamp.
In the work, the entire parameter set of the original model is expressed as $\bm{w}^{(o)}$ while $\bm{w}^{(o)}_k$ denotes only the parameters of layer $k$.
$\varphi()$ is the activation function, e.g., \textit{hyperbolic tangent} (Tanh) or \textit{rectified linear unit} (ReLU).

\subsection{The Attacking Process}

Causative attack allows an attacker to arbitrarily manipulate the input dataset of the model, such as adding, changing, and removing some of the input data.
As one type of causative attack, we assume that poisoning attack allows the attacker to insert arbitrary data $\bm{x}_p$ with artificial label $\bm{t}_p$ into the training dataset.

Here we assume the attacker has already known the details of the original NN model, including the network structure and the exact weight values.
Although the attacker may not directly modify the model, he may still be able to poison the input training data to compromise the model, i.e., to degrade the model's accuracy.
%Attacker cannot directly modify the target model but leverage the information to generate poisoned input samples.
%More specific, through injecting the poisoned data, attacker attempts to compromise the model to minimize its performance accuracy.

Assuming we already have a method to generate the poisoned data, the poisoning attack process can be formulated as below:
\begin{enumerate}[Step 1:]
	\vspace{-6pt}
	\item Generate the poisoned data $\bm{x}_p$ using the given method;
	\vspace{-6pt}
	\item Inject $\bm{x}_p$ into the original model $\bm{w}^{(o)}$ to produce $\bm{w}^{(p)}$;
	\vspace{-6pt}
	\item Send the normal data $\bm{x}_i$ into $\bm{w}^{(p)}$ to get some useful information, e.g., loss and gradients from different time stamps;
	\vspace{-6pt}
	\item Based on the information obtained in Step 3, update the configurations and parameters of the method of poisoned data generation;
	\vspace{-6pt}
	\item Repeat the procedure from Step 1.
\end{enumerate}

\subsection{The Data Gradient and Direct Gradient Method}
The goal of poisoning attack is to minimize the accuracy of the target model $\bm{w}^{(o)}$ by injecting poisoned data $\bm{x}_p$.
Mathematically, this goal is the same as maximizing the loss of the model w.r.t. the normal data $\bm{x}_i$.
When the target model adopts the \textit{mean square error} (MSE) or \textit{cross entropy} as its loss function, the goal of poisoning attack can be respectively represented by
\begin{equation}
\label{equ:mse}
\max_{\bm{x}_p} L(\bm{x}_p)=\sum_{i = 1}^{n}\frac{1}{2}||\varphi_{\bm{w}}(\bm{x}_i)-\bm{t}_i||^2;~\text{or}
\end{equation}
\begin{equation}
\vspace{-8pt}
\label{equ:crossentropy}
\max_{\bm{x}_p} L(\bm{x}_p)=-\sum_{i = 1}^{n}\sum_{j=1}^{n}\bm{1}\{t==j\}\log p(t==j|\bm{x}_i;\bm{w}),
\end{equation}
where $\bm{t}_i$ is the label corresponding to $\bm{x}_i$.

\begin{figure}[t!]
	\begin{center}
		\centerline{\includegraphics[width=5.5cm]{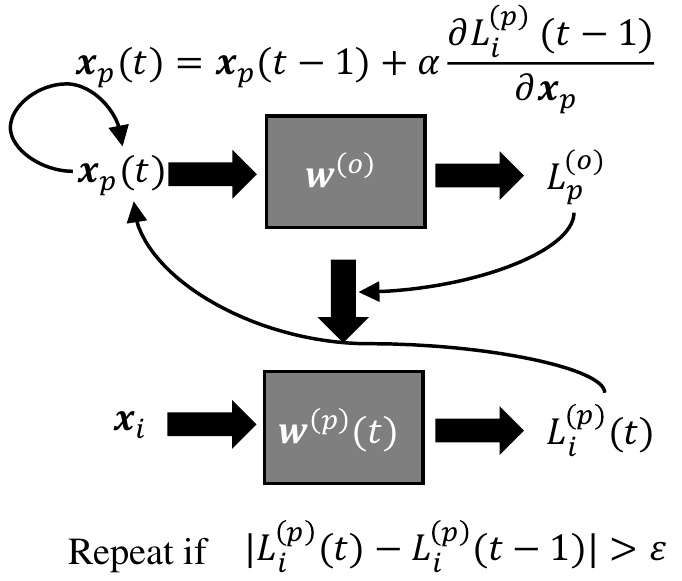}}
		\caption{An overview of direct gradient method.}
		\label{fig:generation}
	\end{center}
\vspace{-16pt}
\end{figure}

As indicated in \textit{Eqs.~(\ref{equ:mse})} and \textit{(\ref{equ:crossentropy})}, the maximum feasible solution in the loss space can be obtained by calculating gradients w.r.t. $\bm{x}_p$ and adopting the gradient ascent procedure.
In this subsection, we will show the mathematical deduction of these gradients in NNs, regardless of the actual loss function. The procedure is depicted in \textit{Figure~\ref{fig:generation}}.
%Here, the back propagation algorithm, matrix derivation, and chain rule are used.

First, we train a NN based on the existing training data $\bm{x}_i$ and obtain the original model $\bm{w}^{(o)}$.
Its corresponding loss $L_i^{(o)}$ w.r.t. $\bm{x}_i$ can be calculated through a forward process on $\bm{w}^{(o)}$, such as
\begin{equation}
\bm{x}_i\xrightarrow{\bm{w}^{(o)}}L_i^{(o)}.
\end{equation}

Next, the poisoned data $\bm{x}_p$ and its corresponding label $\bm{t}_p$ are injected.
There are many ways to choose the initial $\bm{x}_p$, e.g., sampling from a uniform distribution with an arbitrary label or randomly choosing from one of the normal data with an incorrect label (e.g., the label that to be attacked).
We then calculate the loss $L_p^{(o)}$, update the network to $\bm{w}^{(p)}$ through backpropagation with the newly calculated gradients, and obtain the loss $L_i^{(p)}$ as:
\begin{align}
\bm{x}_p&\xrightarrow{\bm{w}^{(o)}}L_p^{(o)},\\
\bm{w}^{(p)}&=\bm{w}^{(o)}-\alpha \frac{\partial L_p^{(o)}}{\partial \bm{w}^{(o)}},~\text{and}\\
\bm{x}_i&\xrightarrow{\bm{w}^{(p)}}L_i^{(p)}.
\end{align}

Finally, we need to calculate the derivative of the sum of $L_i^{(p)}$ w.r.t. $\bm{x}_p$, $\sum\frac{\partial L_i^{(p)}}{\partial \bm{x}_p}$, and then update the poisoned data along the direction of the gradient using a pre-determined coefficient $\bm{x}_p^{\prime}=\bm{x}_p+\alpha\sum\frac{\partial L_i^{(p)}}{\partial \bm{x}_p}$. Here $\frac{\partial L_i^{(p)}}{\partial \bm{x}_p}$ can be expanded using the chain rule and expressed as the multiplication of multiple derivatives, such as
\begin{align}
\label{equ:gradient}
\frac{\partial L_i^{(p)}}{\partial \bm{x}_p}&=\sum_{k = 1}^{m}\frac{\partial L_i^{(p)}}{\partial \bm{w}_k^{(p)}}\frac{\partial \bm{w}_k^{(p)}}{\partial \bm{x}_p},\\
\label{equ:first_derivative}
\frac{\partial L_i^{(p)}}{\partial \bm{w}_k^{(p)}}&={\bm{o}_{k-1}^{(i)}}^\mathrm{T}\bm{\delta}_{k}^{(i)},\\
\label{equ:second_derivative}
\frac{\partial \bm{w}_k^{(p)}}{\partial \bm{x}_p}&=\frac{\bm{w}^{(o)}-\alpha \frac{\partial L_p^{(o)}}{\partial \bm{w}^{(o)}}}{\partial \bm{x}_p}=-\alpha \frac{\partial^2 L_p^{(o)}}{\partial\bm{w}^{(o)}\partial \bm{x}_p},~\text{and}
\end{align}
\begin{equation}
\bm{\delta}_k^{(i)}=\begin{cases}
(\bm{o}_k^{(i)}-\bm{t}_k)\circ\bm{o}_k^{(i)}\circ(1-\bm{o}_k^{(i)})\text{ if $k$ is an output layer}\\
\bm{w}_{k+1}^{(p)}\bm{\delta}_{k+1}^{(i)}\circ\bm{o}_{k}^{(i)}\circ(1-\bm{o}_{k}^{(i)})\text{ if $k$ is an inner layer}.\\
\label{equ:sensitivity}
\end{cases}
\end{equation}
Here $\bm{\delta}_k^{(i)}$ represents the sensitivity of the neurons in layer $k$ to the loss.
$\bm{o}_k^{(i)}=\varphi(\bm{w}_k\bm{o}_{k-1})$ represents the output of the neurons in layer $k$.
$\circ$ is the Hadamard product.

The term $\frac{\partial^2 L_p^{(o)}}{\partial\bm{w}^{(o)}\partial\bm{x}_p}$ in \textit{Eq.~(\ref{equ:second_derivative})} is a second partial derivative.
However, it it difficult to calculate this term with an explicit formula or a backpropagation-alike method because both the first derivatives $\frac{\partial L_p^{(o)}}{\partial\bm{w}^{(o)}}$ and $\frac{\partial L_p^{(o)}}{\partial\bm{x}_p}$ are complex functions of each other.

\begin{algorithm}[htb]
	\caption{Direct gradient method.}
	\label{alg:direct}
	\small
	\begin{algorithmic}
		\STATE \textbf{Input:} Training dataset $D_t$, validation dataset $D_v$, attacking class $\bm{t}_p$, generating rate $\alpha$, threshold of loss $L_{th}$.
		\STATE \textbf{Preprocess:}
		\STATE \textbf{1.} Initialize and train the network $\bm{w}^{(o)}$ with normal training data $\bm{x}_i$ in $D_t$.
		Choose an initial poisoned data from $D_t$ as $\bm{x}_p(0)$ and change its label to $T_a$.
		Reset the round number $t=0$ and the initial loss $\sum L_i^{(p)}(0)=0$.
		\REPEAT
		\STATE \textbf{2.} Inject $\bm{x}_p(t)$ to $\bm{w}^{(o)}$ and update the network to $\bm{w}^{(p)}$;
		\STATE \textbf{3.} Input $\bm{x}_i$ to $\bm{w}^{(p)}$, compute the sum of loss $\sum E_i^{(p)}(t)$;\\
		\STATE \textbf{4.}
		\FOR{$k = 0$; $k<$length of $\bm{x}_p(t)$; $k++$}
		\STATE {Add a small amount $\Delta x$ to $\bm{x}_{p(t)}(k)$};
		\STATE {Inject $\bm{x}_{p(t)}(k)$ to $\bm{w}^{(p)}$ and calculate $L_i^{(p)}(k)$};
		\STATE {Calculate $\frac{L_i^{(p)}}{\bm{x}_p}(k)=\frac{L_i^{(p)}(\bm{x}_{p(t)}(k)+\Delta\bm{x})-L_i^{(p)}(\bm{x}_{p(t)}(k))}{\Delta\bm{x}_{p(t)}(k)}$ };
		\ENDFOR
		\STATE \textbf{5.} Update $\bm{x}_p(t+1) = \bm{x}_p(t)+\alpha\cdot sign(\sum\frac{L_i^{(p)}}{\bm{x}_p}(k))$;
		\STATE \textbf{6.} $t=t+1$;
		\UNTIL $|\sum L_i^{(p)}(t)-\sum L_i^{(p)}(t-1)|<L_{th}$
		\STATE \textbf{Output:} Poisoned data $\bm{x}_p(t)$ with label $\bm{t}_p$.
	\end{algorithmic}
\end{algorithm}

Instead of calculating from \textit{Eq.~(\ref{equ:gradient})-(\ref{equ:sensitivity})}, we propose to directly compute $\frac{L_i^{(p)}}{\bm{x}_p}=\frac{L_i^{(p)}(\bm{x}_p+\Delta\bm{x}_p)-L_i^{(p)}(\bm{x}_p)}{\Delta\bm{x}_p}$ based on its definition.
\textit{Algorithm~\ref{alg:direct}} summarizes the poisoned data generation process of the direct gradient method that is based on our proposed $\frac{L_i^{(p)}}{\bm{x}_p}=\frac{L_i^{(p)}(\bm{x}_p+\Delta\bm{x}_p)-L_i^{(p)}(\bm{x}_p)}{\Delta\bm{x}_p}$ calculation approach.
Here the attacker is provided with a training dataset to generate the target model and the poisoned data.
The attacker is also provided with a validation dataset for testing.
%At the beginning of the process, the initial poisoned data is injected into the target model which will be affected and updated accordingly.
The selection of the initial poisoned data will be discussed in Section~\ref{sec:experiment} in details.
The key of our proposed algorithm is to respectively calculate the gradient w.r.t. each element of $\bm{x}_p$, which will be summed up to a combined gradient $\frac{L_i^{(p)}}{\bm{x}_p}$.
As a result, $\bm{x}_p$ is updated by multiplying the direction of the gradient, i.e., $sign(\frac{L_i^{(p)}}{\bm{x}_p})$, with a coefficient $\alpha$.
The change rate $\alpha$ can be fixed at a pre-determined value (e.g., $\alpha=0.1$) or adaptively change in the process.

As we shall present in Section~\ref{sec:experiment}, the poisoned data generated by direct gradient method can effectively degrade the accuracy of the target model.
However, the data generation can be very time-consuming because of the element-wise gradient calculation.
Very importantly, the computation cost of the poisoned data generation in direct gradient method is proportional to the dimension of the input data and the complexity of the target model.
Hence, the scalability of direct gradient method may be a severe issue, especially considering the rapid growth of the sizes of the popular NNs and their training datasets.

\subsection{A Generative Method to Accelerate Poisoned Data Generation}
\begin{figure}[b]
\vspace{-14pt}
	\begin{center}
		\centerline{\includegraphics[width=6.5cm]{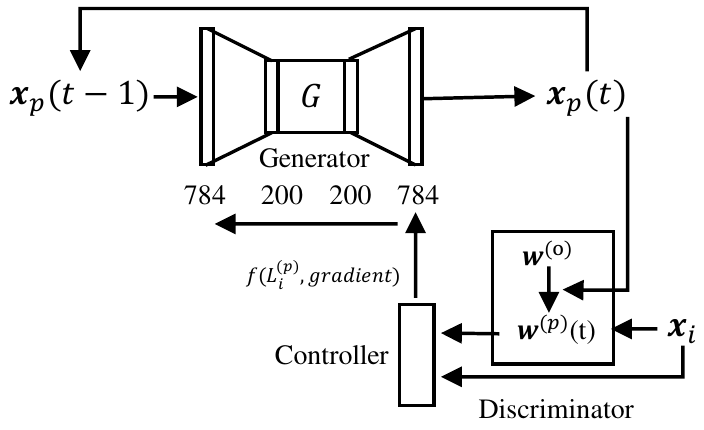}}
		\caption{An overview of the generative method.}
		\label{fig:system}
	\end{center}
\vspace{-12pt}
\end{figure}
Inspired by the concept of GAN~\cite{szegedy2013intriguing}, we propose a generative method which can bypass the costly gradient calculation and therefore speed up the poisoned data generation.% without losing much accuracy degradation.
The key of our method is to train an extra model to generate $\bm{x}_p$.
The gradients of the target model (the first derivative derivation) are sent to the extra model and update the model (another first derivative derivation) in order to generate the new $\bm{x}_p$.
Therefore, the second partial derivative is implicitly calculated.
We name the original target model as \textit{discriminator} ($D$) and the extra model as \textit{generator} ($G$).
In this work, we choose a general autoencoder as our generator.
As a feature extractor, the autoencoder understands the fundamental ingredient of the original pattern, thus changes it from a smaller (compressed) dimension compared with the original one.
The gradients and loss from the discriminator will be collected and calculated to guide the training of the generator.
\textit{Figure~\ref{fig:system}} illustrates the overview of the proposed generative method for poisoned data generation.
The generator is used to generate poisoned data and updated by a weighted function of the loss and the gradients, and then sends the poisoned data to the discriminator.
The discriminator receives the poisoned data and calculates the loss w.r.t. the normal data, and then sends the calculated gradients back to the generator.
The structure of the generator is designed as $784$-$200$-$200$-$784$ for MNIST dataset, which will be adopted in some experiments in Section~\ref{sec:experiment}.

\textit{Algorithm~\ref{alg:generative}} formulates the detailed steps of the generative method.
We assume the same prerequisites of the direct gradient method are provided for the attacker, except for a pre-trained autoencoder serving as a generator. The initial data is sent into the generator in the first step to generate poisoned data, which is then injected to the target model.
The gradients of the target model w.r.t. the normal data $\sum\frac{\partial L_i^{(p)}}{\partial \bm{x}_i}$ are collected, and the obtained weighted gradients are used to update the generator.
In this way, we no longer have to calculate the direct gradients $\frac{L_i^{(p)}}{\bm{x}_p}$ element-wisely and only one update of target model is needed in each iteration. As a result, the time consumption of poisoned data generation is greatly reduced, especially when the target model is complex.

\begin{algorithm}[htb]
	\caption{Generative method.}
	\label{alg:generative}
	\small
	\begin{algorithmic}
		\STATE \textbf{Input:} Training dataset $D_t$, validation dataset $D_v$, attacking class $\bm{t}_p$, generating rate $\alpha$, threshold of loss $L_{th}$.
		\STATE \textbf{Preprocess:}
		\STATE \textbf{1.} Initialize and train the network $\bm{w}^{(o)}$ with normal training data $\bm{x}_i$ in $D_t$.
		Choose a starting poisoned data from $D_t$ as $\bm{x}_p(0)$ and change its label to $T_a$.
		Train an autoencoder-based generator $G$ to generate $\bm{x}_p$.
		Set the round number $t=0$ and the initial loss $\sum L_i^{(p)}(0)=0$.
		\REPEAT
		\STATE \textbf{2.} Input $\bm{x}_p(t)$ to $\bm{w}^{(o)}$, update the network to $\bm{w}^{(p)}$;
		\STATE \textbf{3.} Input $\bm{x}_i$ to updated network $\bm{w}^{(p)}$, compute the sum of loss $\sum L_i^{(p)}(t)$ and the gradients $\sum\frac{\partial L_i^{(p)}}{\partial \bm{x}_i}$;
		\STATE \textbf{4.} Update $G$ with the weighted gradient $f(\sum\frac{\partial L_i^{(p)}}{\partial \bm{x}_i}, \sum L_i^{(p)}(t), \sum L_i^{(p)}(t-1))$;
		\STATE \textbf{5.} Update the poisoned data $\bm{x}_p(t)$ to $\bm{x}_p(t+1)$ via $G$;
		\STATE \textbf{6.} $t=t+1$;
		\UNTIL $\sum L_i^{(p)}(t)-\sum L_i^{(p)}(t-1)<L_{th}$;
		\STATE \textbf{Output:} Poisoned data $\bm{x}_p(t)$ with label $\bm{t}_p$.
	\end{algorithmic}
\end{algorithm}

%The key to this technique of generating poisoned data is how to design the gradients for the generator.
One critical problem here is how to design the gradients for the generator.
We propose to use the difference of two losses from two consecutive attacks to build a reward function.
%Together with the gradients of the target model w.r.t. the normal data, we can obtain a weighted gradient for the generator.
Combing this difference with the gradients of the target model w.r.t. the normal data, we can obtain a weighted gradient for the generator.
%The reward function suggests if this update is a good one (if not, it will be punished), and the gradients of the target model determines the detailed directions of the weighted gradient.
The reward function suggests if this update is a good one while the gradients of the target model determines the detailed directions of the weighted gradient.
If the update is not a good one, it will be punished.
%There could be alternative ways of doing so besides the proposed method, for example, we can only update the encoder part of the generator, by sending back the gradients w.r.t. the output layer of the target model.
%As a result, the number of neurons of the encoder's output layer and the target model's final layer has to be the same.
%Another possible solution is that we could add a policy layer to the generator, as we do in reinforcement learning~\cite{sutton1999policy}.

\subsection{The Loss-based Countermeasure against Poisoning Attack}
Moreover, we propose an universal method to detect the aforementioned poisoning attack methods, which is shown in \textit{Algorithm~\ref{alg:detection}}.
Once a data (no matter normal or poisoned) is injected into the target model, the loss of the target model is recorded.
If the loss exceeds the pre-determined threshold $L_{th}$, a warning will show up.
If the number of warnings exceeds the threshold $W_{th}$, the accuracy check will be triggered to examine if a poisoning attack is indeed being conducted.
\begin{algorithm}[htb]
	\caption{Loss-based poisoning attack detection.}
	\label{alg:detection}
	\small
	\begin{algorithmic}
		\STATE \textbf{Input:} Training dataset $D_t$, validation dataset $D_v$, loss threshold $L_{th}$, warning recorder $R_w=0$, warning threshold $W_{th}$.
		\STATE \textbf{Preprocess:} \textbf{1.} Train the network with $D_t$.
		\WHILE{true}
		\STATE \textbf{2.} Send in training data $\bm{x}$ (no matter normal or poisoned);
		\STATE \textbf{3.} Forward process, calculate the loss $L$;
		\IF {$L>L_{th}$}
		\STATE \textbf{4.} Record a warning, $R_w=R_w+1$;
		\ENDIF
		\IF {$R_w>W_{th}$}
		\STATE \textbf{5.} Calculate the accuracy with validation dataset $D_v$;
		\IF {it is abnormal}
		\STATE \textbf{6.} Set attack alarm to true, $AA=1$, break;
		\ELSE
		\STATE \textbf{7.} Nothing is wrong, clear the warnings, $R_w=0$;
		\ENDIF
		\ENDIF
		\ENDWHILE
		\STATE \textbf{Output:}Attack alarm $AA$.
	\end{algorithmic}
\end{algorithm}
Our method is based on the phenomenon that the poisoned input and label pair usually results in a larger loss, compared with the normal one.
This is naturally understandable as the goal of poisoning attack is to minimize the model accuracy, which is realized through maximizing the loss.
The poisoned data alters target model's original decision boundary.
On the contrary, normal data stays inside the decision region (at least most of time), inducing a relatively smaller loss.
We are able to monitor the condition of inputs by checking the loss periodically. This approach, hence, is much computational less extensively.
%By checking the loss periodically (note that this is easier than checking the validation accuracy which needs a lot of resources for calculation), we can keep monitoring the condition of inputs.

\section{Experiment Results}
\label{sec:experiment}
\begin{figure*}[t!]
	\centering
	\subfigure[Start from normal data ``$5$'' by applying the direct gradient method.]{
		\begin{minipage}{16cm}
			\centering
			\includegraphics[width=16cm]{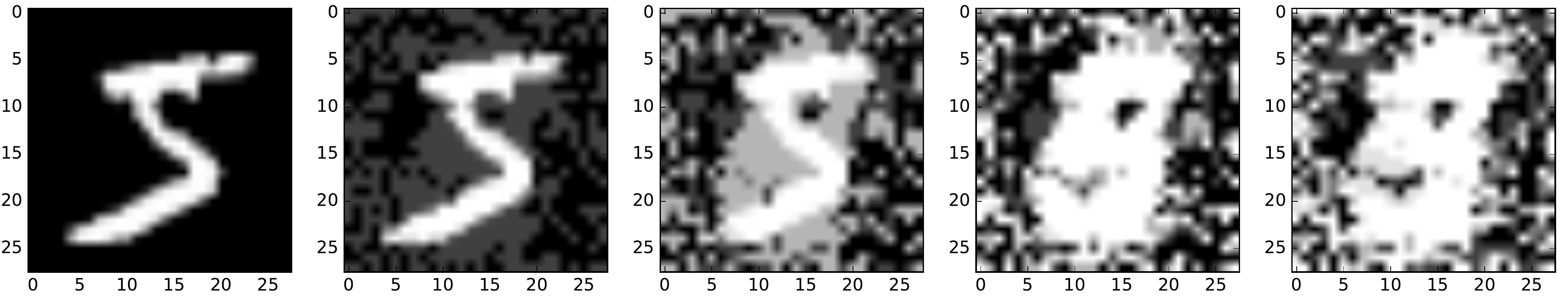} 
		\end{minipage}
	}
	\subfigure[Start from a uniform distribution sampling by applying the direct gradient method.]{
		\begin{minipage}{16cm}
			\centering
			\includegraphics[width=16cm]{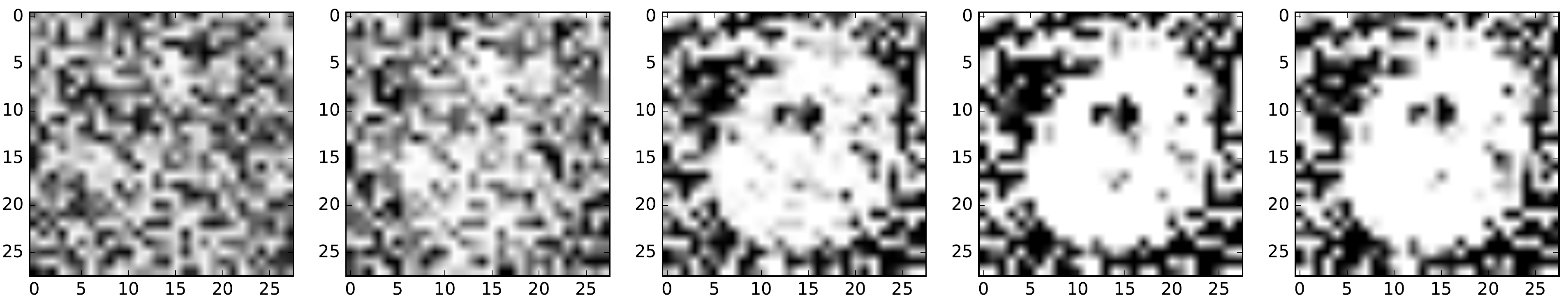} 
		\end{minipage}
	}
	\subfigure[Start from normal data ``$5$'' by applying the generative method.]{
		\begin{minipage}{16cm}
			\centering
			\includegraphics[width=16cm]{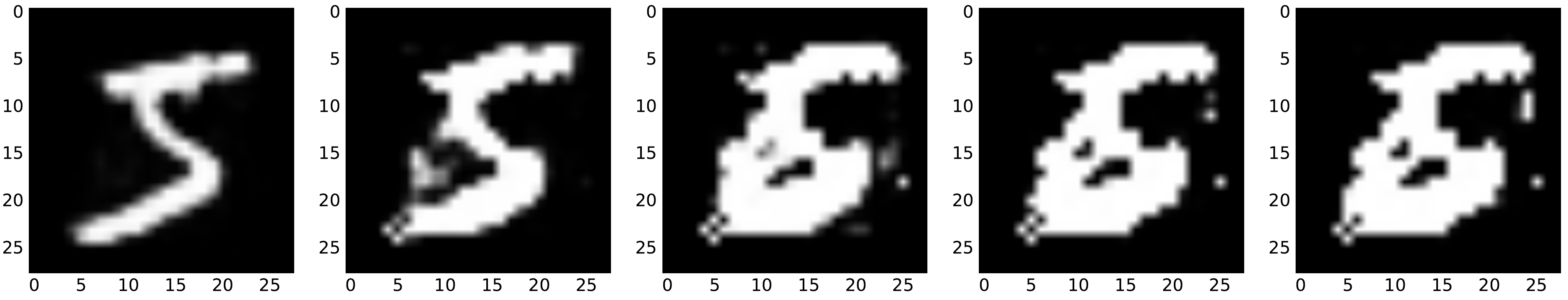} 
		\end{minipage}
	}
	\subfigure[Start from normal data ``bird'' by applying the generative method.]{
		\begin{minipage}{16cm}
			\centering
			\includegraphics[width=16cm]{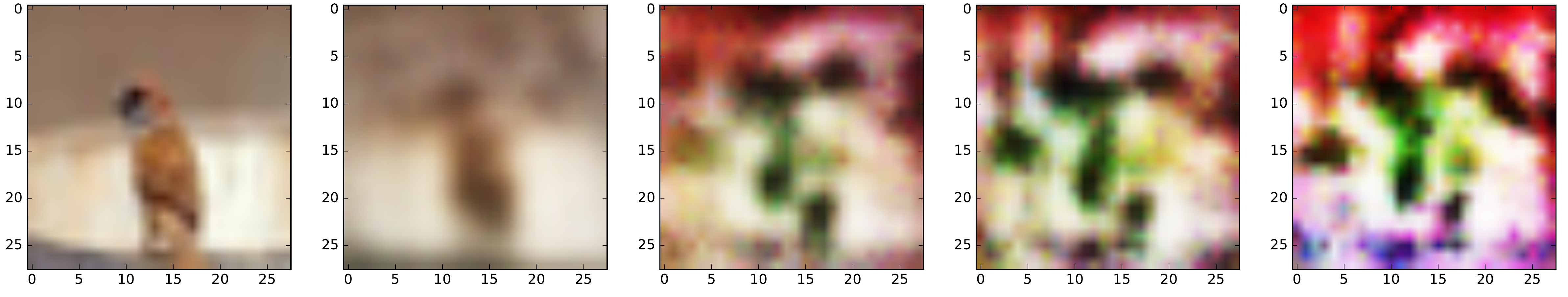} 
		\end{minipage}
	}
	\caption{The process of poisoned data generation under different configurations.}
	\label{fig:poisoned_data}
	\vspace{-12pt}
\end{figure*}
We examine the poisoned data generation process on two datasets -- MNIST~\cite{lecun1998gradient} and CIFAR-10~\cite{krizhevsky2009learning}. 
We compare the target model accuracy degradation and the time consumption of the direct gradient and the generative attack methods.
We also evaluate the feasibility of the loss-based countermeasure technique against poisoning attack.
All the experiments are performed on GeForce GTX 1080 GPU platform.

%\subsection{Effectiveness of Proposed Methods and Poisoned Data Generation }
\subsection{The Effectiveness of Poisoning Attacks Against NN}

Two widely used datasets, MNIST and CIFAR-10, are used to evaluate the effectiveness of the proposed poisoning attack method. 
We choose MXNet~\cite{chen2015mxnet} as our deep learning library. 
The target model for MNIST is a two-layer feed-forward neural network with a structure of $784$-$64$-$10$.
Its original recognition accuracy is $96.82\%$.
We use \textit{Lenet}~\cite{lecun1998gradient} that consists of two convolutional layers and two fully-connected layers as the target model for CIFAR-10, of which the original accuracy is $71.20\%$.
We did not stress the fine-tuning when obtaining these original target models, as the phenomenon of the poisoning attack requires continuous model re-training.
To better demonstrate the data generation process, we conduct the single poisoning attack, that is, injecting one poisoned data per attack. 
%These models are not fine-tuned for that they should serve the purpose of re-training during operation. 
%If we use a fine-tuned model, the attack should be less effective since the new training data has less influence on the model.
%In this experiment, we conduct single poisoning attack, i.e., attack with single poisoned data, to better demonstrate the data generation process.
\begin{figure}[t!]
	\centering
	\subfigure[Group size = 1000.]{
		\begin{minipage}{8cm}
			\centering
			\includegraphics[width=8cm]{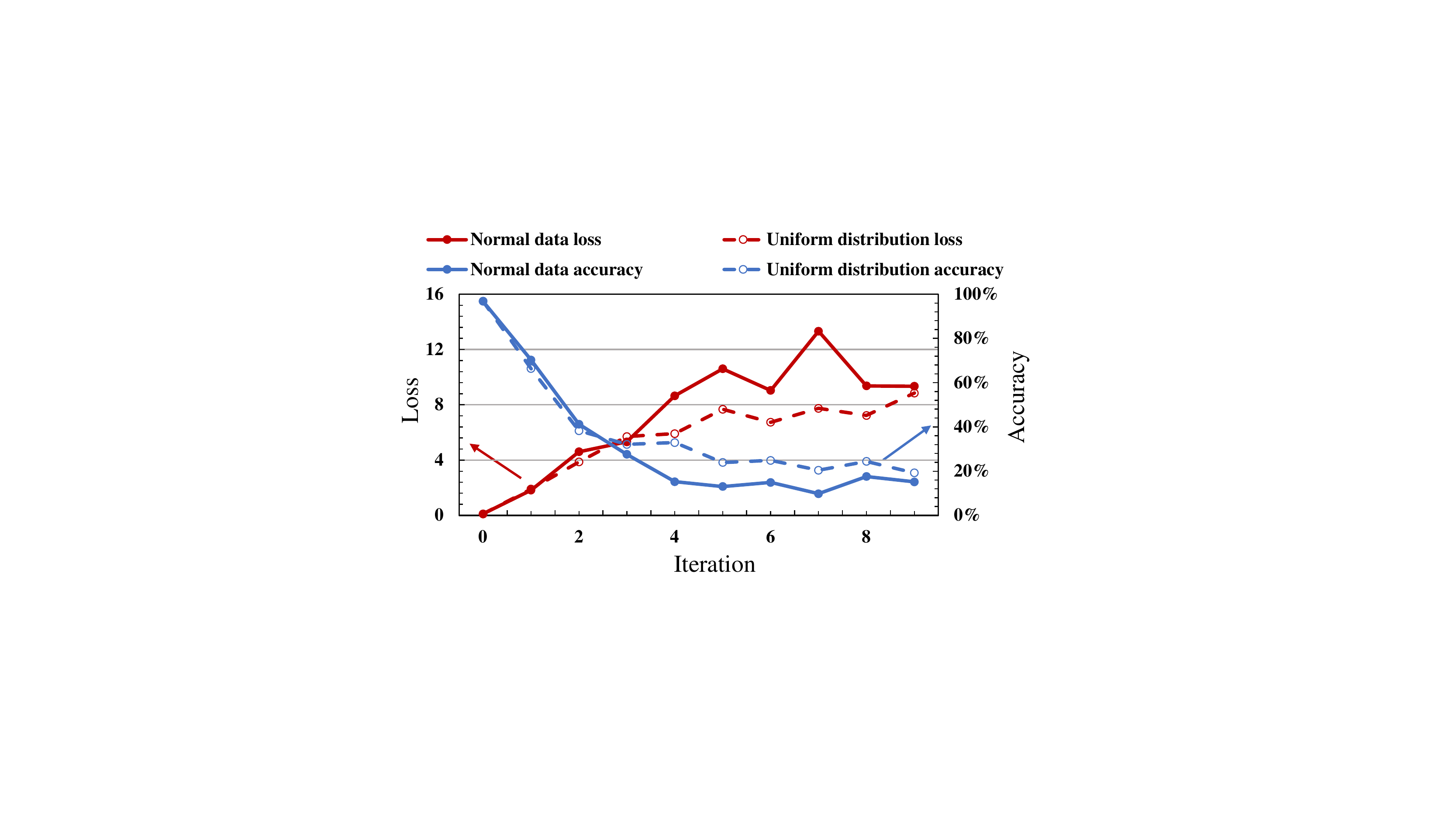} 
		\end{minipage}
	}
	\subfigure[Group size = 100.]{
		\begin{minipage}{8cm}
			\centering
			\includegraphics[width=8cm]{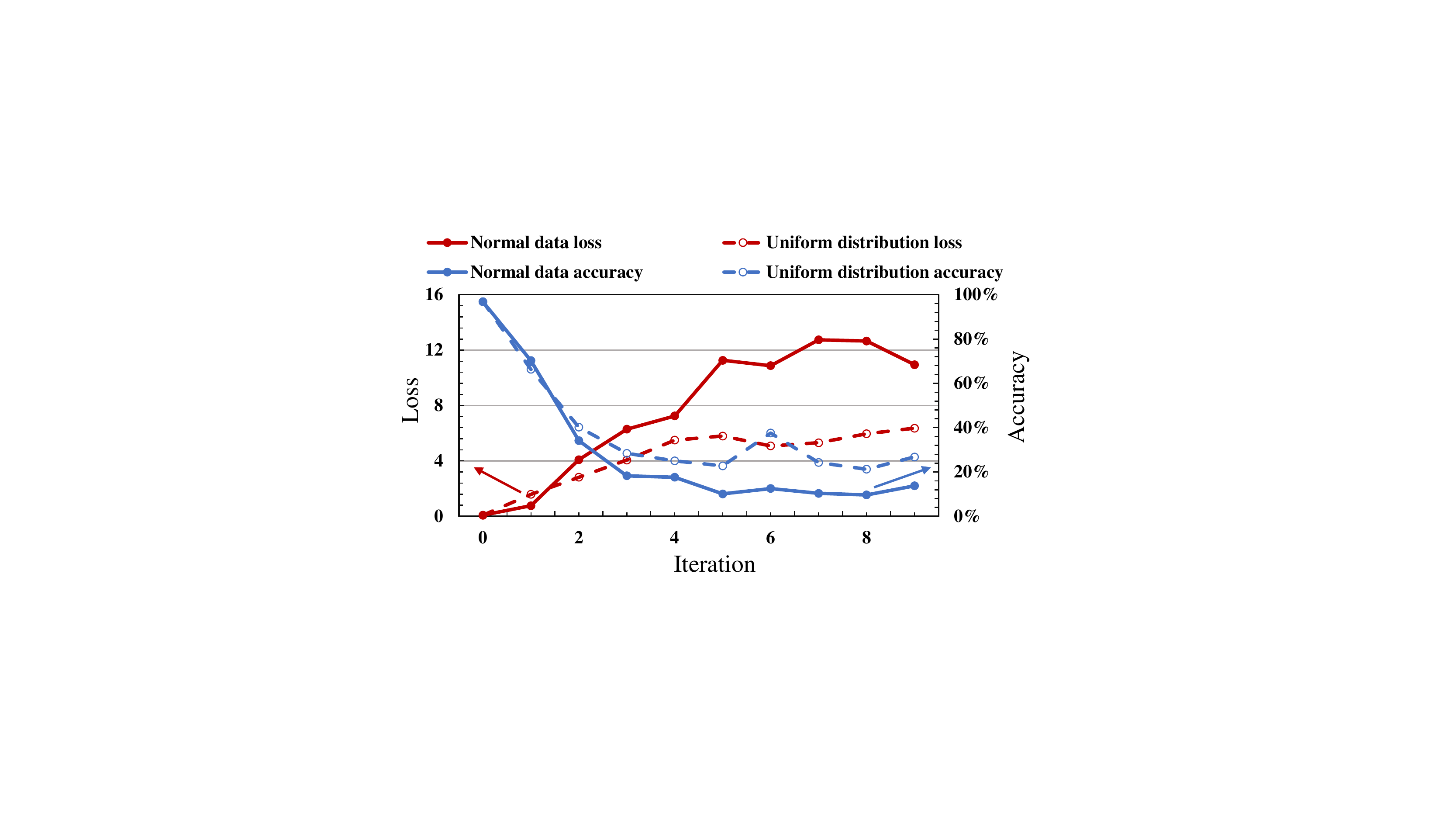} 
		\end{minipage}
	}
	\subfigure[Group size = 10.]{
		\begin{minipage}{8cm}
			\centering
			\includegraphics[width=8cm]{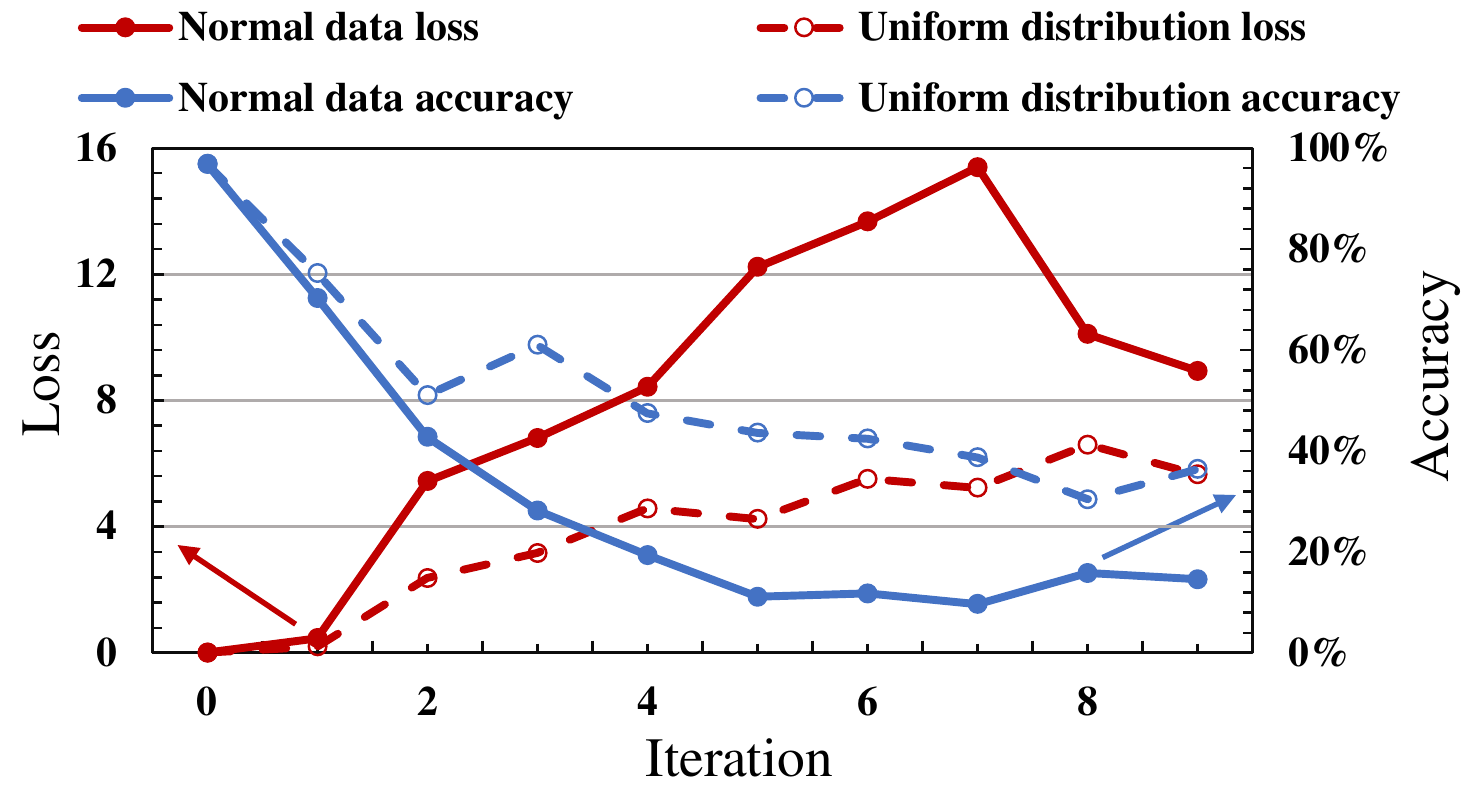} 
		\end{minipage}
	}
	\caption{The trend of the loss and accuracy of the direct gradient method under different group sizes for MNIST dataset.}
	\label{fig:loss_comparison}
	\vspace{-12pt}
\end{figure}

\textit{Figure~\ref{fig:poisoned_data}} demonstrates a few poisoned data generation processes of selected sample images. 
%The first two subfigures apply direct gradient method to MNIST, but start from different initial data. 
%The last two subfigures apply generative method to MNIST and CIFAR-10, respectively, from the normal initial data.
%
Starting from a normal initial data follows the intuitive thinking -- setting the label of a normal data to an incorrect class could compromise the model by making it learn in a wrong way. 
\textit{Figure~\ref{fig:poisoned_data}(a)} applies the direct gradient method to a normal digit $5$ image in MNIST with a wrong label $6$. 
As the impact from gradients accumulates, the generated poisoned image gradually transforms to a pattern alike ``5'' until convergence. 
\textit{Figure~\ref{fig:poisoned_data}(b)} starts from a random data sampling from a uniform distribution with label $6$, where the direct gradient method is used. 
It serves as a control experiment to understand if any preference in the poisoned data generation.
It is interesting that the random data ends as a fuzzy image close to ``$6$''. 
The observation is different from our initial thought that the converged pattern should be very unlike the attacking class, i.e., ``6'' in this example. 
This could possibly be explained as a pattern with a similar structure but very different features to the normal one will compromise the learning process the most.

\textit{Figure~\ref{fig:poisoned_data}(c,d)} demonstrates the poisoned data generation process when applying the generative method. 
For the normal digit $5$ image in MNIST with label $6$ in \textit{Figure~\ref{fig:poisoned_data}(c)}, the generative method shows a similar result as the direct gradient method.
It is hard to identify the exact converged pattern of the \textit{bird} image labeled as \textit{cat} from CIFAR-10.
This is because the data in CIFAR-10 has a much larger dimension than MNIST's (3,072 vs. 784).  

The direct gradient method leverages the loss of the poisoned model w.r.t. the normal data $\sum L_i^{(p)}$. 
Therefore, the group size of the normal data (or, \textit{group size}) that is used to calculate $\sum L_i^{(p)}$ has a strong influence on the convergence speed of the poison data generation and thus the attack effectiveness. 
More specific, large group size provides a more general consideration of the entire dataset, alleviating the impact of some particular data. 
The typical group size is $100$ in this work. 
$10$ and $1000$ are used as references. 
 
\textit{Figure~\ref{fig:loss_comparison}} shows detailed comparisons of the group size and the initial data selection for MNIST dataset.
The blue lines represent the accuracy of the poisoned model and the red line denotes the average loss of the model w.r.t. the normal data group. 
As can be seen from the figures, the loss increase and the accuracy degradation of the normal initial data (solid lines) are much faster than the trend of random initial data (dashed lines). %, especially when the group size is small.
%The normal initial data also converges at a higher loss ($\scriptsize{\sim}11$) and lower accuracy ($\scriptsize{\sim}10\%$).
The results imply that normal initial data is a better choice than the random one. 

In the figure, the first point of a line refers to the result of the original target model, while the second point corresponds to that of the model poisoned by the initial data.
The performance degradation at the second point indicates that normal or random initial data could comprise the target model, but the effectiveness is far from enough. 
For example, the accuracy of the second iteration is $\scriptsize{\sim}70\%$ when the group size is 100. 

Comparing the accuracy degradations in three sub-figures, we note that the curve obtained from a larger group size is more stable than that of a smaller group size. 
When starting from a random data, the larger group size also results in smaller converged accuracy. 
Generally speaking, a larger group size has a better attack effectiveness but also introduces more computation overhead.
Detailed performance comparison shall be presented and discussed in Section~\ref{sec:experiment:comp}.

The normal initial data of the small target model designed for MINIST dataset converges at $\scriptsize{\sim}10\%$. 
In reality, regularly continuous re-training process for a neural network in a much larger scale could alleviate the attack.
So we may not obtain such a significant accuracy degradation in the poisoning attack. 
Even though, the results prove the effectiveness of our proposed methods and suggest potential study prospect. 

%\subsection{Comparison with Different Configurations}
\subsection{Direct Gradient vs. Generative Methods}
\label{sec:experiment:comp}

%\begin{table*}[t]
%	\caption{The comparison of the direct gradient method and the generative method on time, accuracy, and loss.}
%	\label{tab:comparison}
%	\begin{center}
%		\begin{small}
%			\begin{sc}
%				\begin{tabular}{cccccccc}
%					\toprule
%					\multicolumn{2}{c}{Method}&MNIST, G&MNIST, D&MNIST, S&CIFAR-10, G&CIFAR-10, D&CIFAR-10, S\\ 
%					\midrule
%					\multirow{3}{*}{$1000$}&time (s)&$3.88\pm0.68$&$345.85\pm10.50$&$89.14\times $&$14.40\pm1.14$&$3447.8\pm83.55$&$239.38\times$\\
%					&accuracy ($\%$)&$16.59$&$\bm{8.84}$&$0.53\times$&$20.74$&$\bm{20.51}$&$0.99\times$\\
%					&loss&$13.67$&$13.32$&$0.97\times$&$\bm{7.34}$&$4.37$&$0.60\times$\\
%					\hline
%					\multirow{3}{*}{$100$}&time (s)&$\bm{3.41\pm0.98}$&$36.18\pm3.13$&$10.61\times$&$\bm{11.69\pm1.46}$&$580.70\pm17.44$&$49.67\times$\\
%					&accuracy ($\%$)&$17.03$&$9.64$&$0.57\times$&$20.88$&$20.91$&$1.00\times$\\
%					&loss&$12.37$&$12.97$&$1.05\times$&$5.73$&$4.45$&$0.78\times$\\
%					\hline
%					\multirow{3}{*}{$10$}&time (s)&$3.84\pm0.60$&$4.18\pm0.91$&$1.09\times$&$12.56\pm0.77$&$259.60\pm6.56$&$20.67\times$\\
%					&accuracy ($\%$)&$16.41$&$9.67$&$0.62\times$&$23.40$&$20.81$&$0.89\times$\\
%					&loss&$14.75$&$\bm{15.39}$&$1.04\times$&$5.54$&$5.48$&$0.99\times$\\
%					\bottomrule
%				\end{tabular}
%			\end{sc}
%		\end{small}
%	\end{center}
%	\vspace{-12pt}
%\end{table*}
\begin{table*}[t]
	\caption{The comparison of the direct gradient method and the generative method on time, accuracy, and loss.}
	\label{tab:comparison}
	\begin{center}
		\begin{small}
			\begin{sc}
				\begin{tabular}{cc|ccc|ccc}
					\toprule
					%\multicolumn{2}{c|}{Method}&MNIST,G&MNIST,D&MNIST,S&CIFAR-10,G&CIFAR-10,D&CIFAR-10,S\\ 
					\multicolumn{2}{c|}{}&\multicolumn{3}{c|}{MNIST}&\multicolumn{3}{c}{CIFAR-10}\\
					\multicolumn{2}{c|}{}&Generative&Gradient&Improve&Generative&Gradient&Improve\\
					\midrule
					\multirow{3}{*}{$1000$}&time (s)&$3.88\pm0.68$&$345.85\pm10.50$&$89.14\times $&$14.40\pm1.14$&$3447.8\pm83.55$&$239.38\times$\\
					&accuracy ($\%$)&$16.59$&$\bm{8.84}$&$0.53\times$&$20.74$&$\bm{20.51}$&$0.99\times$\\
					&loss&$13.67$&$13.32$&$0.97\times$&$\bm{7.34}$&$4.37$&$0.60\times$\\
					\midrule
					\multirow{3}{*}{$100$}&time (s)&$\bm{3.41\pm0.98}$&$36.18\pm3.13$&$10.61\times$&$\bm{11.69\pm1.46}$&$580.70\pm17.44$&$49.67\times$\\
					&accuracy ($\%$)&$17.03$&$9.64$&$0.57\times$&$20.88$&$20.91$&$1.00\times$\\
					&loss&$12.37$&$12.97$&$1.05\times$&$5.73$&$4.45$&$0.78\times$\\
					\midrule
					\multirow{3}{*}{$10$}&time (s)&$3.84\pm0.60$&$4.18\pm0.91$&$1.09\times$&$12.56\pm0.77$&$259.60\pm6.56$&$20.67\times$\\
					&accuracy ($\%$)&$16.41$&$9.67$&$0.62\times$&$23.40$&$20.81$&$0.89\times$\\
					&loss&$14.75$&$\bm{15.39}$&$1.04\times$&$5.54$&$5.48$&$0.99\times$\\
					\bottomrule
				\end{tabular}
			\end{sc}
		\end{small}
	\end{center}
	\vspace{-12pt}
\end{table*}

We compare the direct gradient and the generative methods for the two datasets with the normal initial data, under three group size settings (1000, 100, 10).
Here, the structure of the autoencoder used as the generator for MNIST is $784$-$200$-$200$-$784$. 
The generator for CIFAR-10 is a deeper fully-connected autoencoder with 6 layers.
%Other types of autoencoder including the one with transposed convolutional layers~\cite{radford2015unsupervised} are also worth trying.
\textit{Table~\ref{tab:comparison}} summarizes the comparison in terms of the average time to generate one poisoned data ({\sc Time}), the least accuracy achieved by the target model ({\sc Accuracy}), and the largest average normal data loss achieved of the data group  ({\sc Loss}).
The improvement of the generative method over the direct gradient method (faster time, lower accuracy, and higher loss) are also presented in the table. 
The best result of each metric under each dataset is marked as bold.

According to the table, the direct gradient method costs more time than the generative method especially as the group size increases.
For MNIST dataset, the generative method converges at a higher accuracy (i.e., the worst case is $16.59\%$ vs. $8.84\%$) which is still in a reasonable range. 
This phenomenon is due to the fact that we use a small network on a small dataset, and the accuracy degradation converges too fast for generative method.
For the bigger dataset CIFAR-10, the two methods show the similar accuracy levels.
It implies that the generative method has a better scalability and works better for larger neural networks. 
Moreover, the time cost of the direct gradient method has a strong dependency on the group size while it is not observed in the generative method.
In summary, the generative method is more effective and expects to have a great potential in poisoning attacking larger neural network models and bigger datasets. 

\subsection{Evaluating The Proposed Poisoning Attack Detection}
\begin{figure}[t!]
	\begin{center}
		\centerline{\includegraphics[width=8cm]{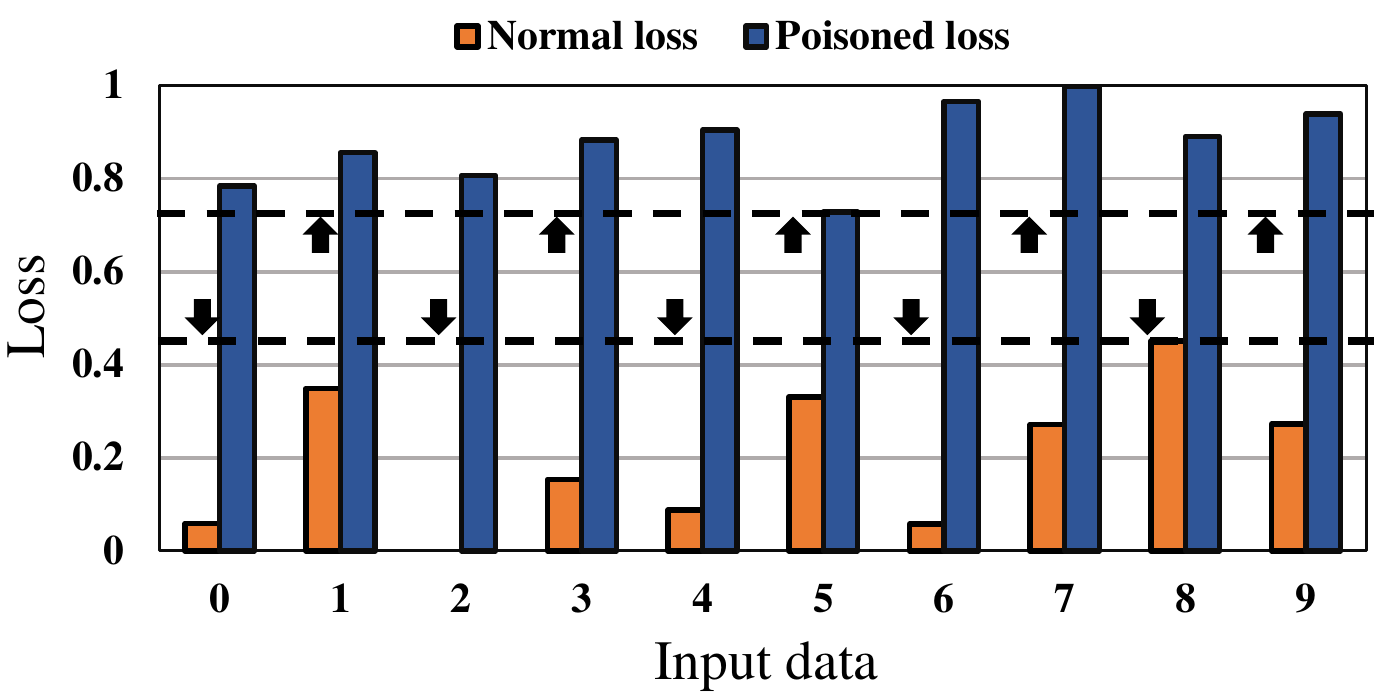}}
		\caption{The loss difference between normal and poisoned losses.}
		\label{fig:detection}
	\end{center}
\vspace{-36pt}
\end{figure}
As mentioned in Section~\ref{sec:algorithm}, poisoned data deviates the decision boundary of the target model and aggravate its loss. 
Hence, periodically monitoring the loss anomaly appears to be an effective way of detecting poisoning attack.
For the MNIST dataset, we measure the difference between the normal loss and the loss introduced by poisoning attack (or, poisoned loss) and show the result in \textit{Figure~\ref{fig:detection}}. 
The horizontal axis represents different randomly selected input data with artificial labels. 
The vertical axis denotes the normalized loss incurred by re-training the target model on these data. 
Each point on the vertical axis is bind with the above two losses. 
A clear gap exists between the least poisoned loss and the largest normal loss. 
This is even true if we choose the input data from a uniform distribution %(right five pairs of the losses in the figure), 
which is the worst case without human interference. 
The result proves the feasibility of our proposed poisoning attack detection method.

Moreover, the loss of the target model is usually calculated during re-training. 
Implementing the proposed poisoning attack detection technique in \textit{Algorithm~\ref{alg:detection}}, therefore, requires marginal extra computation and hardware overheads.
%This fact is especially crucial for neuromorphic computing chips with a great need for low-power design. 
%We only need to design a mechanism, e.g., the one proposed in Section~\ref{sec:algorithm}, and take advantage of the already calculated loss. 
It is impossible to completely prevent poisoning attack since no one can learn right from wrong. 
However, we could extend the monitor mechanism to other parameters and criteria of the model to provide a fast alarm system which can invoke the poisoning attack detection in time. 

\section{Conclusion}
\label{sec:conclusion}
In this paper, we study the general process of poisoned data generation for neural networks (NNs). 
We propose two poisoning attack methods against NNs, including a direct gradient method and a generative method. 
We also develop a preliminary countermeasure that performs a loss-based poisoning attack detection. 
Experiments on MNIST and CIFAR-10 show that the generative method can substantially improve the poisoned data generation rate compared with the direct gradient method. 
%with less effectiveness on degrading the accuracy of the target model.
Particularly the former one demonstrates great potential in attacking large neural network models and big datasets. 
%promising result on the achieved accuracy. In this work, we demonstrate our proposed method on relatively small datasets networks, for better clarification and trials. Real life problem usually involves much more difficult and complex scenarios.
Our future work will focus on designing a better generator and gradient policy that can achieve both improved poisoned data generation rate and model attack effectiveness. 
%Therefore, Our future work aims to extend current work to complex, dynamic, and real-time scenario. We also seek to design a better generator and gradient policy, for a high-speed, effective generation method.

\section*{Acknowledgment}

The presented works were supported by NSF CNS-1253424, CCF-1615475 and AFRL FA8750-15-2-0048.

\bibliographystyle{abbrv}
\bibliography{ref}

\end{document}